\documentclass[runningheads,a4paper]{llncs}

\usepackage{amssymb}
\usepackage{graphicx}

\usepackage[text={15.5cm,23cm},centering]{geometry}
\usepackage{color}
\usepackage[colorlinks=true, linkcolor=blue]{hyperref}
\usepackage[version=4]{mhchem} % chem package
\usepackage[flushleft]{threeparttable}
\usepackage{multirow}

\usepackage{url}
\urldef{\mailsa}\path|pgao@carnegiescience.edu|

\begin{document}

\mainmatter  % start of an individual contribution

\title{Clouds and Hazes in the Atmospheres of Triton and Pluto}

\titlerunning{Clouds and Hazes on Triton and Pluto}

\author{Peter Gao$^1$ \and Kazumasa Ohno$^2$}

%\thanks{Please note that the LNCS Editorial assumes that all authors have used the western naming convention, with given names preceding surnames. This determines the structure of the names in the running heads and the author index.}
%
\authorrunning{Gao \& Ohno}

\institute{$^1$Earth \& Planets Laboratory, Carnegie Institution for Science, Washington, DC, USA\\
$^2$Division of Science, National Astronomical Observatory of Japan, Tokyo, Japan \\
\mailsa}

%\mailsa
%\url{http://www.springer.com/lncs}}


%\toctitle{Lecture Notes in Computer Science}
%\tocauthor{Authors' Instructions}
\maketitle

%\begin{abstract}
%The abstract should summarize the contents of the paper and should
%contain at least 70 and at most 150 words. It should be written using the
%\emph{abstract} environment.
%\keywords{We would like to encourage you to list your keywords within
%the abstract section}
%\end{abstract}

\section{Introduction}

Clouds and hazes are abundant in the thin ($\sim$10 $\mu$bar) and cold ($\leq$40 K) atmospheres of Triton and Pluto, where they are thought to be produced by interactions between atmospheric gases and ultraviolet (UV) photons from the Sun and those scattered by the local interstellar medium. These interactions lead to a rich network of chemical reactions that produces higher order hydrocarbons and nitriles that condense out to form ice clouds, and ultimately complex haze particles that rain down onto the surface that impact the atmospheric thermal structure, gas chemistry, and surface evolution. In this chapter, we will review the observational evidence for clouds and hazes in the atmospheres of Triton and Pluto and theoretical interpretations thereof, and the emerging set of experiments aiming to produce Triton and Pluto clouds and hazes in the lab to learn about them in detail. 

\subsection{A Note on Nomenclature}\label{sec:nomen}

It is important to give formal definitions to ``clouds'' and ``hazes'', as their definitions vary between Earth and planetary science literature. Here we will follow the latter and use definitions related to provenance: \textit{Cloud} particles form through equilibrium condensation reactions, where the local temperature and condensate vapor abundance are such that condensation is energetically favorable and occurs spontaneously. In contrast, \textit{haze} particles form through disequilibrium processes, typically photochemical reactions, where injection of energy (e.g. UV photons) breaks apart atmospheric gas molecules and leads to reactions that gradually build the haze particles. As a result, cloud particles are typically volatile while haze particles are typically not. A complication of these definitions involves particles that form from condensation of gases that were themselves produced photochemically; while we will refer to these structures as clouds here, they have also been referred to as hazes in the literature. Finally, we will use ``aerosols'' as a catch-all term for any particulates suspended in an atmosphere, i.e. both clouds and hazes\footnote{This is also different from its definition in Earth literature.}, in cases where the provenance is uncertain. 

\section{Observations of Triton and Pluto Aerosols}\label{sec:obs}

\subsection{Voyager 2 Observations of Triton Aerosols}\label{sec:tritonvoy2}

\begin{figure}[t]
\begin{center}
\includegraphics[width=\textwidth]{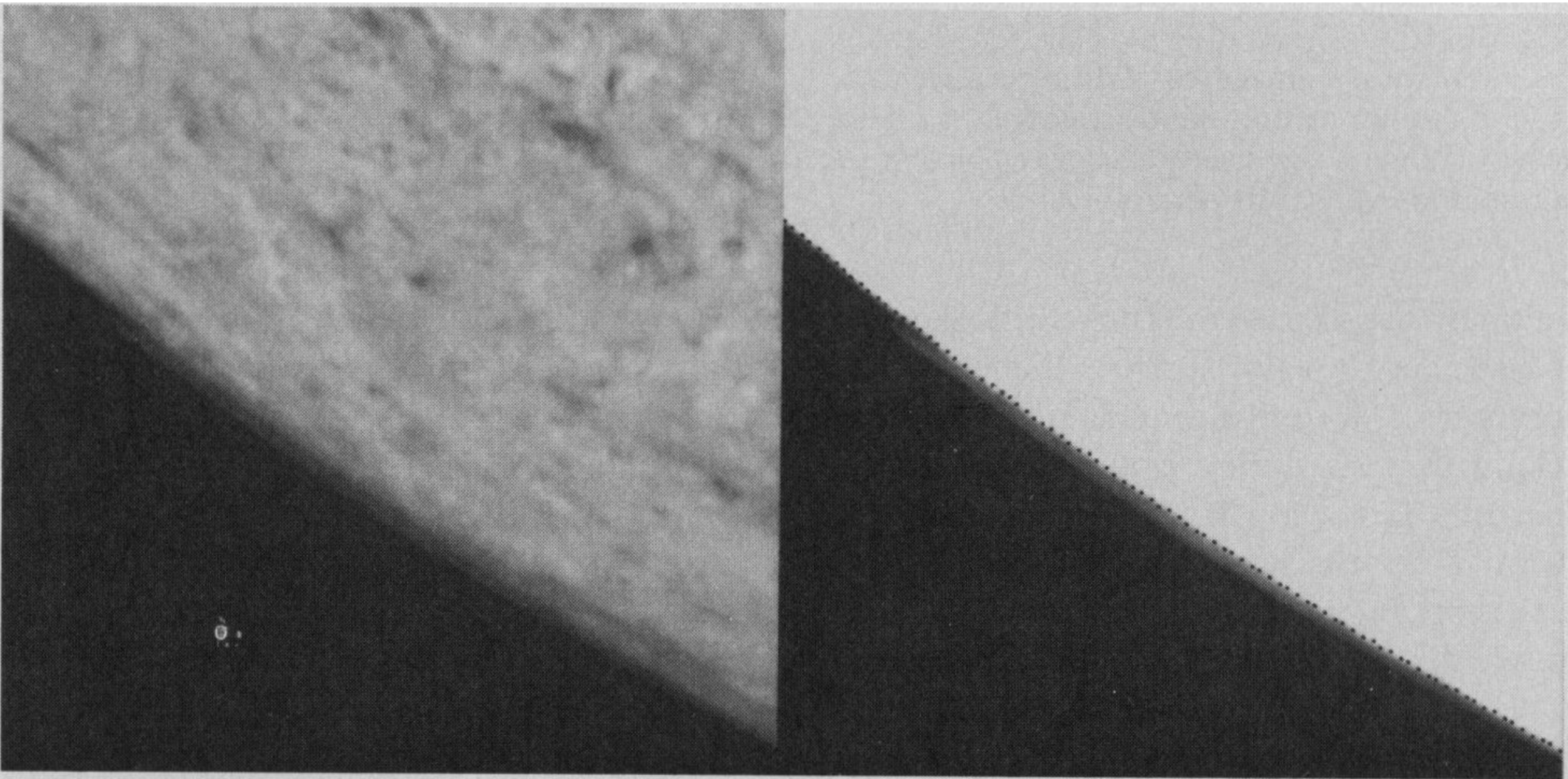}
\end{center}
\vspace{-5mm}
\caption{Images of aerosols over Triton's south polar cap taken at the west limb. The left panel highlights the surface features of Triton, while the right panel highlights the limb aerosols, with dark pixels marking Triton's limb. Figure taken from \cite{Smith1989Sci...246.1422S}.}
\label{fig:image_Triton}
\end{figure}

Voyager 2's historic flyby of the Neptune system on 25 August 1989 revealed the existence of a \ce{N2}-dominated atmosphere on Triton with surface pressure and temperature of 16 $\pm$ 3 $\mu$bar and 38$^{+3}_{-4}$ K, respectively, consistent with conditions of vapor equilibrium with surface \ce{N2} ice \cite{Broadfoot1989Sci...246.1459B,Tyler1989Sci...246.1466T,Conrath1989Sci...246.1454C}. The observed near-surface mixing ratio of \ce{CH4} (few to a few hundred ppm) was subsaturated by a factor of 30 and spatially variable with a scale height of 7-10 km, which is considerably smaller than expected from hydrostatic equilibrium and thus suggested the action of photochemistry \cite{Broadfoot1989Sci...246.1459B}. Imaging observations uncovered two aerosol components: bright, discrete structures at $8~{\rm km}$ and mostly poleward of 30$^{\circ}$S (Figure \ref{fig:image_Triton}), and a global aerosol layer that extends to $\sim30~{\rm km}$ at almost everywhere on Triton \cite{Smith1989Sci...246.1422S,Pollack1990Sci...250..440P}.

\begin{figure}[t!]
\begin{center}
\includegraphics[width=0.7\textwidth]{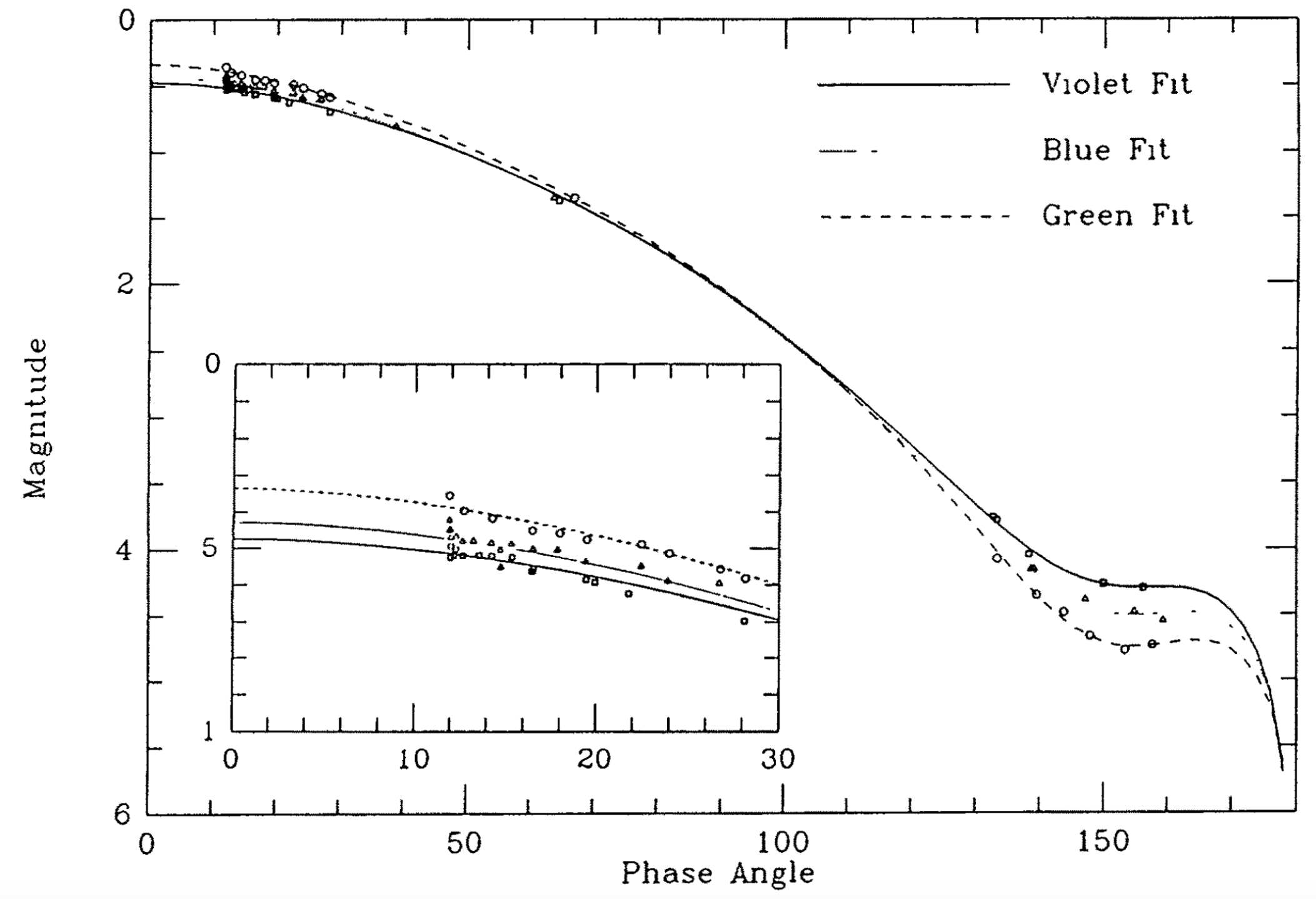}
\end{center}
\vspace{-5mm}
\caption{Model phase curves of Triton \cite{Hillier1990Sci...250..419H} compared with the Voyager 2 violet, blue, and green filter data.  The vertical axis is normalized such that the magnitude at the zero degree phase is $-2.5\log{(p_{\rm geo})}$, where $p_{\rm geo}$ is the geometric albedo. The disk-averaged image is brighter at smaller phase angles due to surface scattering, while the upturn at large phase angles is due to the scattered light from aerosols. Figure is taken from \cite{Hillier1990Sci...250..419H}.}
\label{fig:IF_Triton}
\end{figure}

Voyager 2 revealed the scattering properties of Triton's aerosols with photometric observations at visible wavelengths (Figure \ref{fig:IF_Triton}). The upturn in the scattered light brightness at large phase angles strongly indicates the presence of reflective aerosols. From the disk-averaged brightness, which predominantly traces the photometric properties of the near-surface discrete structures \cite{Hillier1994Icar..109..284H}, the vertical aerosol optical depth was constrained to $\sim0.03$ at $\lambda=0.56~{\rm {\mu}m}$ and followed a wavelength dependence of $\propto \lambda^{-2}$, with an asymmetry parameter $\sim0.6$ (highly forward scattering) and a single scattering albedo $\sim0.99$ \cite{Hillier1990Sci...250..419H,Hillier1991JGR....9619203H}. However, it should be noted that the single scattering albedo cannot be independently determined apart from the aerosol optical depth for optically thin aerosols.

Spatially-resolved observations help to avoid the ambiguity between the discrete and global aerosol structures when constraining their scattering properties. From spatially-resolved images at high phase angles, the average particle size of the discrete structures at high southern latitudes was constrained to be $\sim0.2$--$1.5~{\rm {\mu}m}$ \cite{Pollack1990Sci...250..440P,Rages1992Icar...99..289R}. Meanwhile, at low southern latitudes where the discrete structures are largely absent, the mean particle radius, column-integrated particle number density, and aerosol scale height of the global aerosol layer were estimated to be $0.17\pm0.012~{\rm {\mu}m}$, $2.0\pm0.6\times{10}^{6}~{\rm {cm}^{-2}}$, and $11\pm0.6~{\rm km}$, respectively, yielding vertical scattering optical depths of $\sim$0.002-0.004 across the visible wavelengths \cite{Rages1992Icar...99..289R}. The fact that the derived scale height was considerably smaller than the pressure scale height at the observed region ($\sim 16~{\rm km}$) suggests that aerosol formation occurred at altitudes below 20 km, as the aerosol and pressure scale heights would be identical otherwise. 

The vertical extinction profile of Triton's global aerosol layer was constrained by stellar occultation observations from Voyager 2's UV spectrometer. The observed extinction at $0.14$--$0.16~{\rm {\mu}m}$ at altitudes $<20~{\rm km}$ had a wavelength dependence suggestive of submicron ($<$0.3 ${\rm {\mu}m}$) particles, as the inferred optical depth at UV wavelengths was orders of magnitude higher than that at visible wavelengths derived from the imaging observations \cite{Smith1989Sci...246.1422S,Pollack1990Sci...250..440P}. The extinction also differed between ingress and egress \cite{Herbert1991JGR....9619241H}, while the extinction scale height was identical to the pressure scale height, in contrast to the smaller scale height obtained from spatially-resolved, visible observations \cite{Rages1992Icar...99..289R}.

\subsection{Pre-New Horizons Observations of Pluto's Atmosphere}\label{sec:plutocc}

\begin{figure}[t]
\begin{center}
\includegraphics[width=0.7\textwidth]{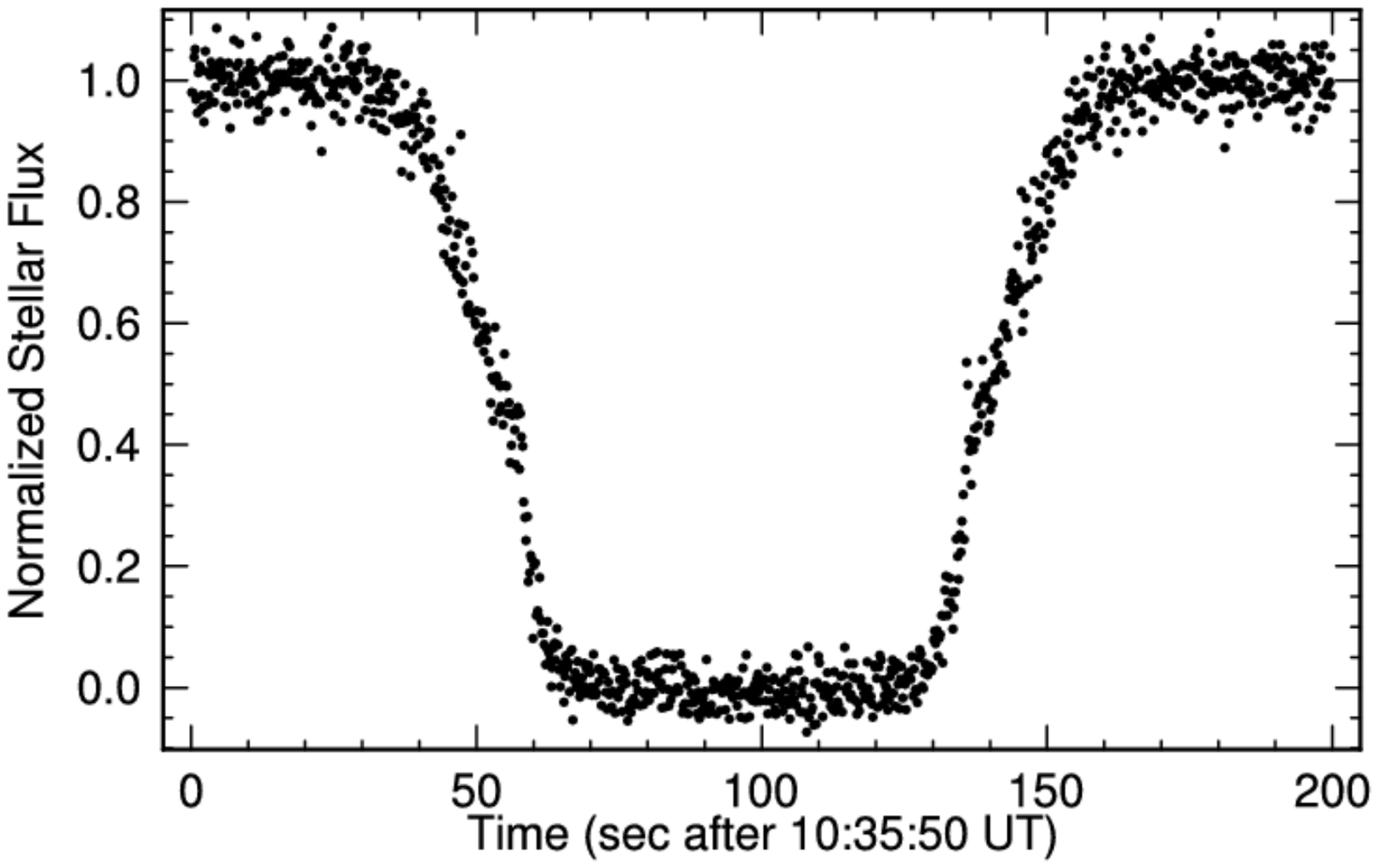}%{pluto_occultation_1988_KAO.pdf}
\end{center}
\vspace{-5mm}
\caption{Light curve of the star P8 from the Pluto occultation of 9 June 1988 observed by the Kuiper Airborne Observatory. Figure obtained from \cite{Elliot2003AJ....126.1041E}. }%(points) fit with an isothermal atmospheric model that includes refraction and extinction (solid curve) from \cite{Elliot1992AJ....103..991E}. The dashed curve shows the light curve if extinction were not included. The pluses at the bottom show the fit residuals to the well-fit model. Figure obtained from \cite{Elliot1996AREPS..24...89E}. }
\label{fig:lightcurve}
\end{figure}

Pluto's atmosphere was detected from groundbased observations of a stellar occultation on 9 June 1988 \cite{Hubbard1988Natur.336..452H,Elliot1989Icar...77..148E,Millis1993Icar..105..282M}, which showed a gradual -- rather than sudden, in the case of a bare surface -- loss of flux from the target star as Pluto occulted it (Figure \ref{fig:lightcurve}). The ``knee'' in the light curve at $\sim$55 seconds after 10:35:50 UTC, where the stellar flux drops more precipitously, was the first hint that aerosols may pervade the lower atmosphere of Pluto. This is because extinction by aerosols would reduce the stellar flux more quickly than a clear isothermal atmosphere \cite{Elliot1992AJ....103..991E}. However, the ``knee'' can also be explained by a reduction in the atmospheric scale height due to a drastic temperature decrease near the surface \cite{Hubbard1990Icar...84....1H}. Subsequent stellar occultations from 2002 to 2015, e.g. \cite{Elliot2003Natur.424..165E,Pasachoff2005AJ....129.1718P,Young2008AJ....136.1757Y,Person2008AJ....136.1510P,Olkin2014Icar..239...15O,Gulbis2015Icar..246..226G} found evidence for atmospheric waves and showed that both aerosols and a steep temperature gradient were likely present. Evidence for aerosol extinction was provided by the wavelength-dependence of minimum stellar fluxes consistent with a fall off in aerosol extinction caused by the decrease in absorption of submicron aerosol particles at longer wavelengths \cite{Elliot2003Natur.424..165E,Gulbis2015Icar..246..226G}. 

The atmospheric composition of Pluto was initially inferred from detections of \ce{N2}, \ce{CH4}, and \ce{CO} ices on its surface via infrared spectroscopy \cite{Cruikshank1980Icar...41...96C,Owen1993Sci...261..745O}, which implied an atmosphere of similar composition in vapor equilibrium. The finding that \ce{N2} ice was more abundant than the others by a factor of 50, and subsequent measurement of the low surface temperature of Pluto of $\sim$40 K \cite{Tryka1994Icar..112..513T} all but confirmed \ce{N2} as the dominant atmospheric gas due to its volatility. Direct detections of gaseous \ce{CH4} and \ce{CO} were subsequently made through high resolution spectroscopy \cite{Young1997Icar..127..258Y,Lellouch2011A&A...530L...4L}. 

\subsection{New Horizons Observations of Pluto Aerosols}\label{sec:plutonh}

\begin{figure}[t!]
\begin{center}
\includegraphics[width=\textwidth]{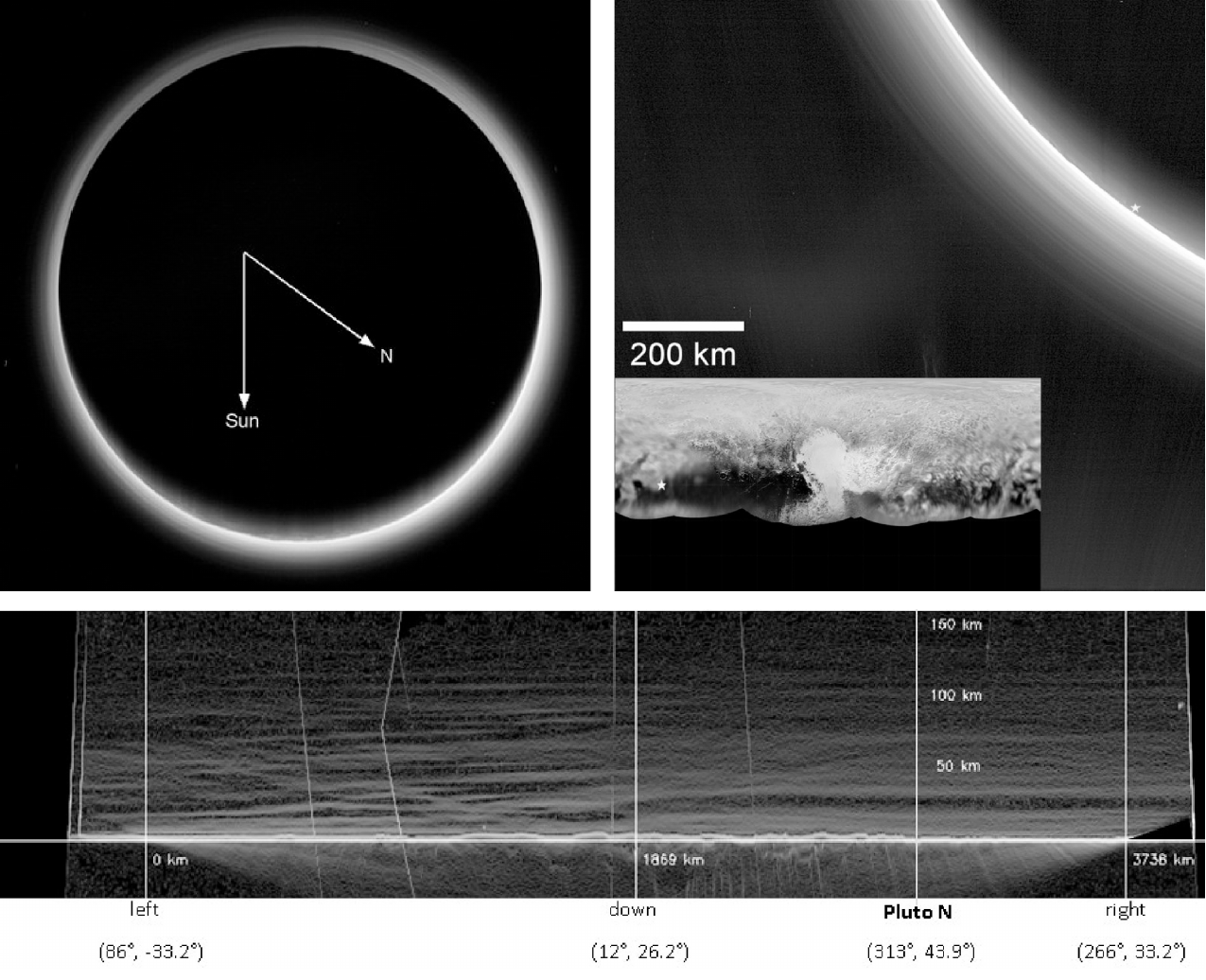}
\end{center}
\vspace{-5mm}
\caption{\textbf{Top Left:} New Horizon's view of Pluto backlit by the Sun, showing the aerosol layers around the full limb. Aerosol scattering is brighter towards the northern (summer) hemisphere rather than the Sun's direction. Image has been stretched and sharpened. \textbf{Top Right:} Close-up of the $\sim$20 aerosol layers at Pluto's limb, with the inset map of Pluto showing the location of the observed aerosols. \textbf{Bottom:} Unwrapped mosaic of haze layers around the Pluto limb. Thin, tilted white lines are mosaic seams. Horizontal distance and vertical altitude scales and Pluto longitude/latitude are given. All images are taken by the New Horizons Long Range
Reconnaissance Imager (LORRI) and obtained from  \cite{Cheng2017Icar..290..112C}. }
\label{fig:nhhaze}
\end{figure}

\begin{figure}[t!]
\begin{center}
\includegraphics[width=1.0\textwidth]{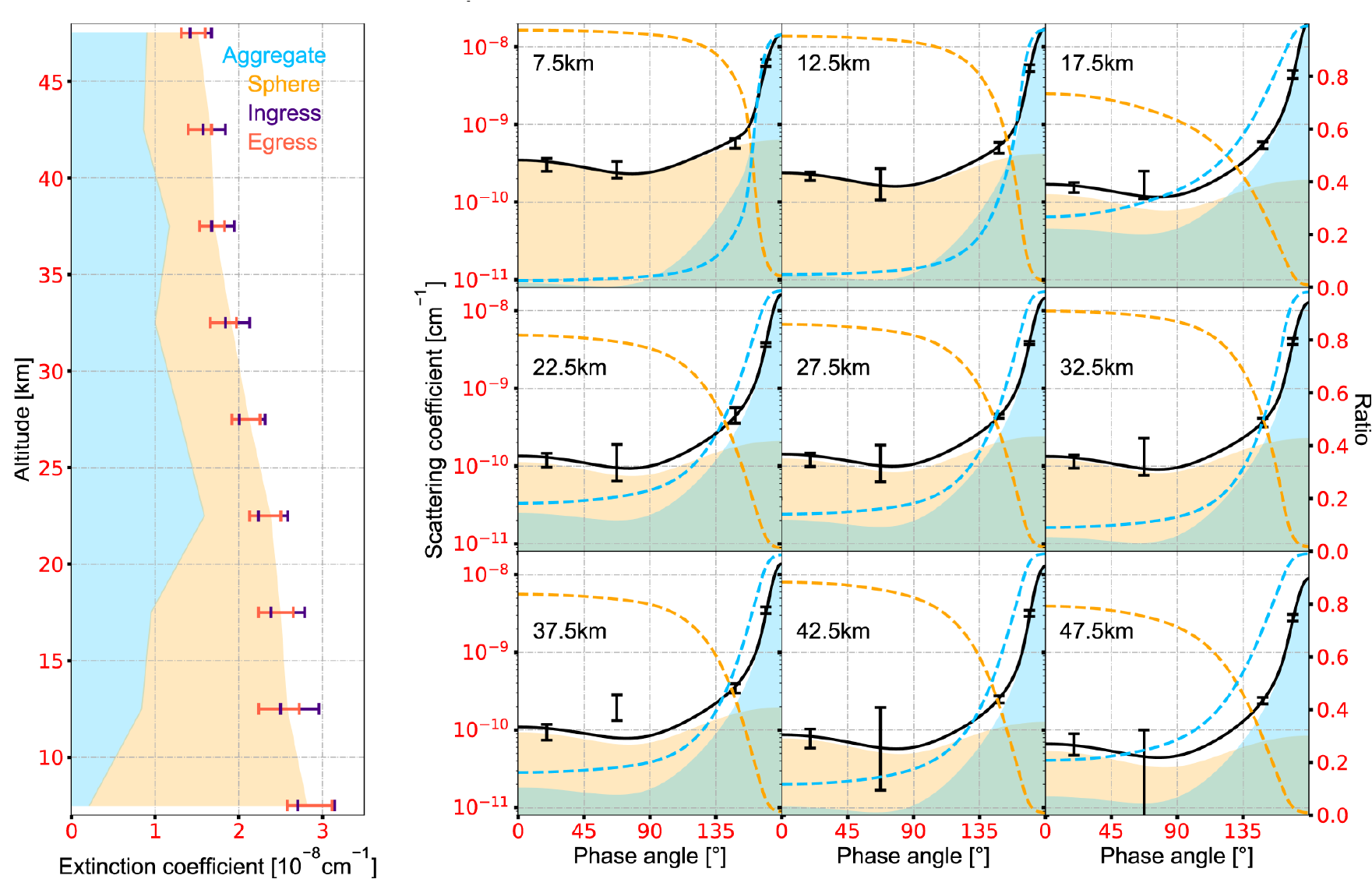}
\end{center}
\vspace{-5mm}
\caption{\textbf{Left:} Near-surface aerosol extinction at Pluto observed by the Alice spectrograph during ingress (indigo) and egress (red) of solar occultation compared to the modeled contributions from large $\sim$1 $\mu$m aggregates (blue) and smaller particles of a few tens of nm radius (orange). \textbf{Right:} Near-surface scattered light observations obtained by LORRI (black) compared to the modeled contributions from large aggregates and small particles. Black curves denote the total contributions of aggregates and spheres. Colored dashed curves represent the ratio of contribution from each component. Figures taken from \cite{Fan2022NatCo..13..240F}. }
\label{fig:phasedep}
\end{figure}

The New Horizons spacecraft flew through the Pluto system on 14 July 2015 and confirmed the existence of aerosols in Pluto's atmosphere (Figure \ref{fig:nhhaze}). The aerosols were found to be optically thin but global, with nadir optical depths of $\sim$0.01 at optical wavelengths but wrapping around the full limb \cite{Gladstone2016Sci...351.8866G}. Unlike Triton, no discrete aerosol structures were detected conclusively \cite{Stern2017AJ....154...43S}. Aerosol scattering was brighter towards the northern (summer) hemisphere by a factor of 2 to 3 compared to equatorial regions, suggesting a factor of 2 greater aerosol mass loading. Aerosol scattering at optical wavelengths was observed to 200 km altitude above the surface, while UV extinction by aerosols was seen reaching 500 km. The vertical distribution of aerosols was not smooth, but consisted of $\sim$20 fine distinct layers each with a thickness of a few km separated on average by about 10 km, though in general aerosol scattering increased with decreasing altitude. Contrast between the bright layers and the darker regions in between is a few percent. Each thin aerosol layer was observed to extend for hundreds of km around the limb, and sometimes merging with others or splitting apart \cite{Cheng2017Icar..290..112C,Jacobs2021Icar..35613825J}. Pluto's aerosols appear to be more Titan-like and absorbing than Triton's, with single scattering albedo ranging from 0.9-0.95 over wavelengths of 500 to 900 nm  \cite{Hillier2021PSJ.....2...11H}. 

The wavelength and phase angle dependence of aerosol extinction point to a complex aerosol particle distribution (Figure \ref{fig:phasedep}). The aerosol scattering intensity in the visible and near-IR showed a gradual decrease with increasing wavelength, suggesting aerosol particle sizes of $\sim$10 nm \cite{Gladstone2016Sci...351.8866G,Grundy2018Icar..314..232G,Kutsop2021PSJ.....2...91K}. However, the aerosols were also found to be highly forward scattering in the visible, which is indicative of larger ($>$0.1 $\mu$m) particles. Furthermore, the ultraviolet aerosol extinction indicated a large aerosol UV cross section that cannot be produced by even 0.1 $\mu$m spherical particles \cite{Young2018Icar..300..174Y,Kammer2020AJ....159...26K}. Porous fractal aggregates -- irregularly shaped, extended particles composed of loosely bound, small, roughly spherical monomers (see ${\S}$\ref{sec:micro}) -- provide a possible explanation for these seemingly conflicting observations \cite{West1991Icar...90..330W}: the wavelength dependence and large UV cross section are due to scattering and extinction by the collection of $\sim$10 nm monomers, while the forward scattering is accomplished by the large ($>$0.1 $\mu$m) extended nature of the particles \cite{Cheng2017Icar..290..112C}. However, while fractal aggregate particles fit the observations at altitudes above 50 km, an increase in the aerosol backscattering intensity near the surface cannot be reproduced by this population alone. Here, two separate populations of aerosol particles are required: smaller particles (spherical or aggregate) with radii of a few tens of nm that are capable of producing the observed backscattering, and $\sim$1 $\mu$m fractal aggregates with $\sim$10 nm monomers that can explain the large forward scattering and UV extinction and wavelength dependence (Figure \ref{fig:phasedep}) \cite{Kutsop2021PSJ.....2...91K,Fan2022NatCo..13..240F}. 

The abundance profiles of several key gases in Pluto's atmosphere were constrained through solar and stellar UV occultations observed by the New Horizons Alice UV spectrograph  \cite{Young2018Icar..300..174Y,Kammer2020AJ....159...26K}. \ce{N2} and \ce{CH4} were shown to be the dominant gases in the atmosphere, with mixing ratios of 99\% and 1\% near the surface, respectively. Diffusive separation at high altitudes results in the mixing ratio of \ce{CH4} approaching 5\% at $\sim$500 km. \ce{C2H2}, \ce{C2H4}, and \ce{C2H6} were also identified, which are likely photochemical in origin (see ${\S}$\ref{sec:photo}). Interestingly, these species are not well-mixed in the atmosphere below their nominal photochemical production altitude, and instead show depletion below 400 km suggestive of chemical destruction and/or condensation. The mixing ratio profile of \ce{HCN}, which was measured by ALMA at nearly the same time as the New Horizons flyby, shows similar decreases below 400 km, but also a highly supersaturated mixing ratio in the upper atmosphere \cite{Lellouch2017Icar..286..289L}.

\subsection{Summary}

Observations of Triton and Pluto's atmospheres show that both worlds possess global aerosol layers with intriguing differences and similarities. The biggest difference is the vertical extent of the aerosols: while Triton's aerosols are confined to the lower $\sim$30 km of the atmosphere, Pluto's extend to $>$500 km. In addition, Triton possesses two distinct aerosol populations: a low optical depth global layer and higher optical depth near-surface discrete structures. In contrast, Pluto lacks discrete structures (though there are extensive vertical layering, which is not present on Triton), but still appears to possess two particle populations near the surface, with particle sizes similar to the two Triton aerosol populations. The total vertical optical depths of the aerosols on both worlds are also strikingly similar at around 0.01, though Triton's is more variable due to the discrete structures.

\section{Theory of Triton and Pluto Aerosols}\label{sec:theory}

\subsection{Photochemistry and Aerosol Formation on Triton and Pluto}\label{sec:photo}

The origins of aerosols in the cold reducing atmospheres of Triton and Pluto are inexorably tied to the photochemistry of \ce{N2} and \ce{CH4}. \ce{N2} is destroyed by extreme-UV (EUV) solar photons ($\lambda<$100 nm) and, additionally for Triton, high energy electrons from Neptune's magnetosphere \cite{Majeed1990GeoRL..17.1721M,Yung1990GeoRL..17.1717Y,Lara1997Icar..130...16L}; \ce{CH4} is photolyzed mostly by Lyman-$\alpha$ photons from the Sun and those scattered by the local interstellar medium  \cite{Strobel1990GeoRL..17.1729S,Lara1997Icar..130...16L}. These reactions yield radical species such as \ce{N}, \ce{H}, \ce{CH}, $^3$\ce{CH2}, and \ce{CH3}, which rapidly react with each other to form more complex hydrocarbons and nitriles like \ce{C2H2}, \ce{C2H4}, \ce{C2H6}, \ce{C4H2}, and \ce{HCN} \cite{Strobel1990GeoRL..17.1729S,Krasnopolsky1995JGR...10021271K,Lara1997Icar..130...16L,Wong2017Icar..287..110W,Luspay-Kuti2017MNRAS.472..104L,Benne2022A&A...667A.169B}.

\begin{figure}[t]
\centering
%\begin{center}
\includegraphics[width=0.7\textwidth]{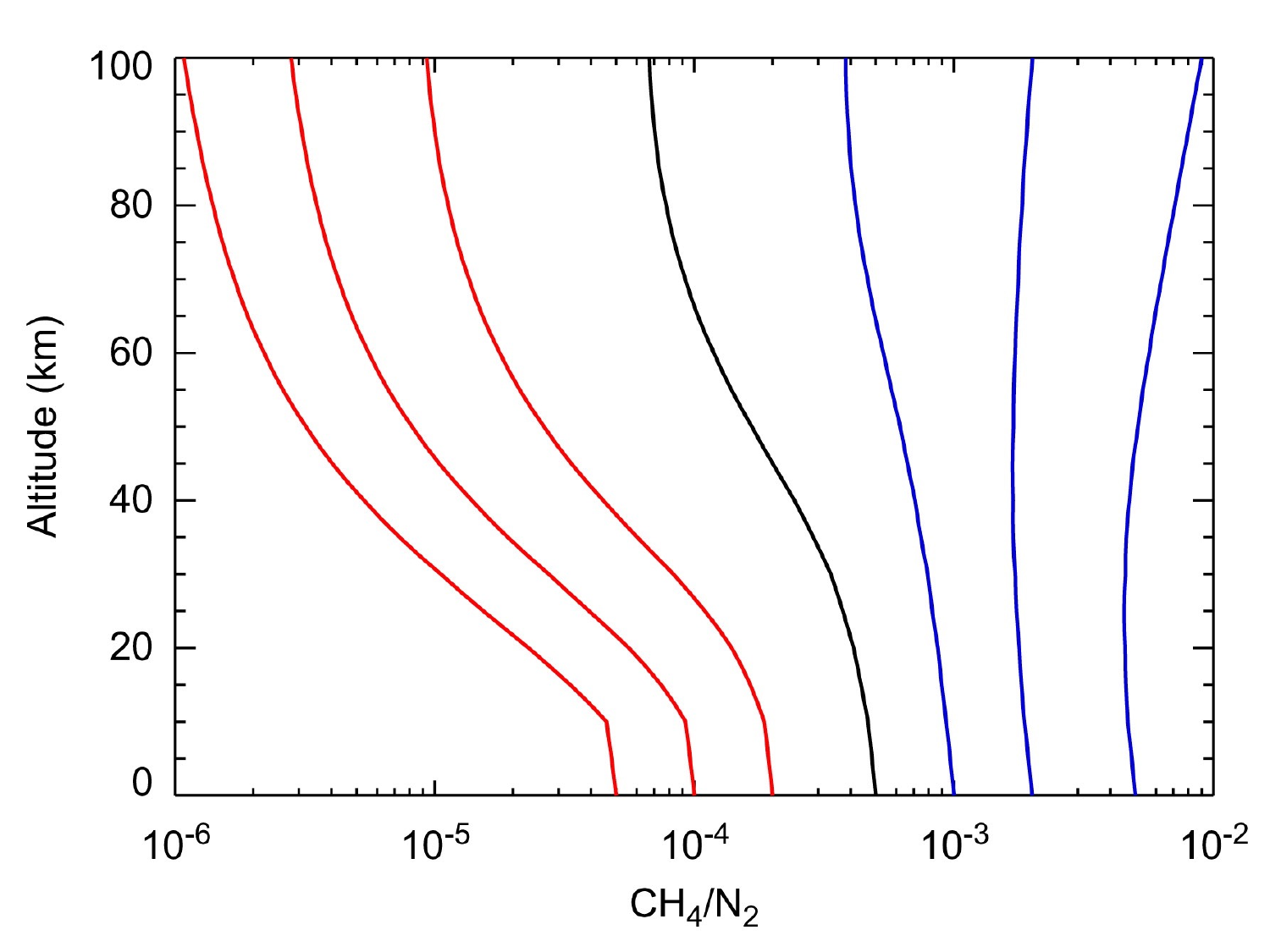}
%\end{center}
\vspace{-5mm}
\caption{\ce{CH4} mixing ratio profiles on Triton as functions of its value near the surface. Photolysis and chemical reactions tend to reduce methane (red) while diffusive separation enhances it (blue). Transition between the two regimes (black) occur at a surface \ce{CH4} mixing ratio of $\sim$5 $\times$ 10$^{-4}$. Figure taken from \cite{Krasnopolsky2012P&SS...73..318K}.}
\label{fig:photochem}
\end{figure}

A key difference in the photochemistry of Triton and Pluto is the near-surface gaseous \ce{CH4} abundance, which is a factor of 10-100 smaller on the former compared to the latter. This results in \ce{CH4} being photolyzed, and thus more complex hydrocarbons and nitriles being produced, at a much lower altitude on Triton -- at around 25 km \cite{Krasnopolsky2012P&SS...73..318K} -- in contrast to 400 km on Pluto \cite{Lara1997Icar..130...16L,Wong2017Icar..287..110W,Krasnopolsky2020Icar..33513374K}. The rapid destruction of \ce{CH4} near the surface of Triton explains the observed small \ce{CH4} scale height of 7-10 km and the lack of \ce{CH4} at higher altitudes \cite{Broadfoot1989Sci...246.1459B}. In contrast, the abundant \ce{CH4} on Pluto overwhelms any photochemical destruction, allowing it to persist into the upper atmosphere (Figure \ref{fig:photochem}) \cite{Krasnopolsky2012P&SS...73..318K,Young2017Icar..284..443Y}.

Lessons learned from the analysis, modeling, and experimental interpretation of Cassini observations of Saturn's moon, Titan, can be applied to understand the connection between photochemistry and haze formation on Triton and Pluto. All three bodies possess \ce{N2}-dominated atmospheres with trace amounts of \ce{CH4}, though Titan's atmosphere is significantly more massive. Similar to Triton and Pluto, \ce{N2} and \ce{CH4} are destroyed in the upper atmosphere and ionosphere of Titan ($\sim$1000 km altitude) through photolysis by EUV and FUV (far-UV) photons and energetic particles from Saturn's magnetosphere \cite{Agren2009P&SS...57.1821A,Galand2010JGRA..115.7312G}, producing radicals and progressively more complex organic molecules and ions, e.g. \cite{Yung1984ApJS...55..465Y,Wilson2004JGRE..109.6002W,Vuitton2019Icar..324..120V}. Cassini observations show that this process of atmospheric species gradually increasing in complexity and mass through successive neutral and ion reactions continues smoothly up to heavy negative ions with masses of $\sim$10$^4$ Da q$^{-1}$ \cite{Coates2007GeoRL..3422103C,Vuitton2009P&SS...57.1558V}. These negative ions attract the less massive (up to 350 Da q$^{-1}$ \cite{Crary2009P&SS...57.1847C}) positive ions in Titan's ionosphere and grow rapidly as a result, eventually forming the nm-sized ``seeds'' of Titan's global hazes \cite{Lavvas2013}. 

The complex chemistry and formation of hazes on Titan depends heavily on the coupling between neutral and ion chemistry in its ionosphere. Titan's ionosphere exhibits an electron density peaking at $\sim$3000 e$^-$ cm$^{-3}$ \cite{Agren2009P&SS...57.1821A} and a smorgasbord of organic molecules and ions sourced from \ce{N2} and \ce{CH4}. In comparison, New Horizons only placed an upper limit of 1000 e$^-$ cm$^{-3}$ on Pluto's ionosphere, which is otherwise populated by positive molecular ions \cite{Hinson2018Icar..307...17H}. Triton's ionosphere exhibits an even greater electron density than Titan of 25000-45000 e$^-$ cm$^{-3}$ \cite{Tyler1989Sci...246.1466T} due to the lack of \ce{CH4} allowing the survival of atomic species and ions \cite{Krasnopolsky1995JGR...10021271K}, which in turn possess long enough lifetimes to ensure high electron densities %(Figure \ref{fig:photochem}) 
\cite{Strobel&Summers95_book}. However, the low abundance of \ce{CH4} at high altitudes also makes complex organic chemistry difficult. In addition, the atmospheres of Triton and Pluto both contain about 10 times the abundance of \ce{CO} as Titan (500 ppm vs. 50 ppm) \cite{Gurwell2004ApJ...616L...7G,Lellouch2010A&A...512L...8L,Lellouch2017Icar..286..289L}. These differences could result in significant disparities in the haze properties of these atmospheres.

\subsection{Formation, Evolution, and Dynamics of Triton and Pluto Aerosols}\label{sec:micro}

The size and spatial distributions of aerosols are controlled by transport processes such as sedimentation, diffusion, and advection, as well as microphysical processes including nucleation, growth through condensation and coagulation, and loss through evaporation/sublimation. In the following sections, we summarize several key microphysical processes relevant to Triton and Pluto aerosols and what models that consider these processes have shown. 

\subsubsection{Haze Particle Transport}

Upon formation, nm-sized nascent haze particles will begin settling downwards at their terminal velocities due to gravity. At the low gas densities of Pluto and Triton's atmosphere, the mean free path of gas molecules is much larger than the haze particle radii (large Knudsen number), so that particle transport occurs in the gas kinetic regime. As a result, the terminal velocity is linearly proportional to the particle mass density and radius (for compact particles) and inversely proportional to the ambient gas density.%, and the flow of gas over the particles remains laminar (low Reynolds number). 

\begin{figure}[t]
\begin{center}
\includegraphics[width=\textwidth]{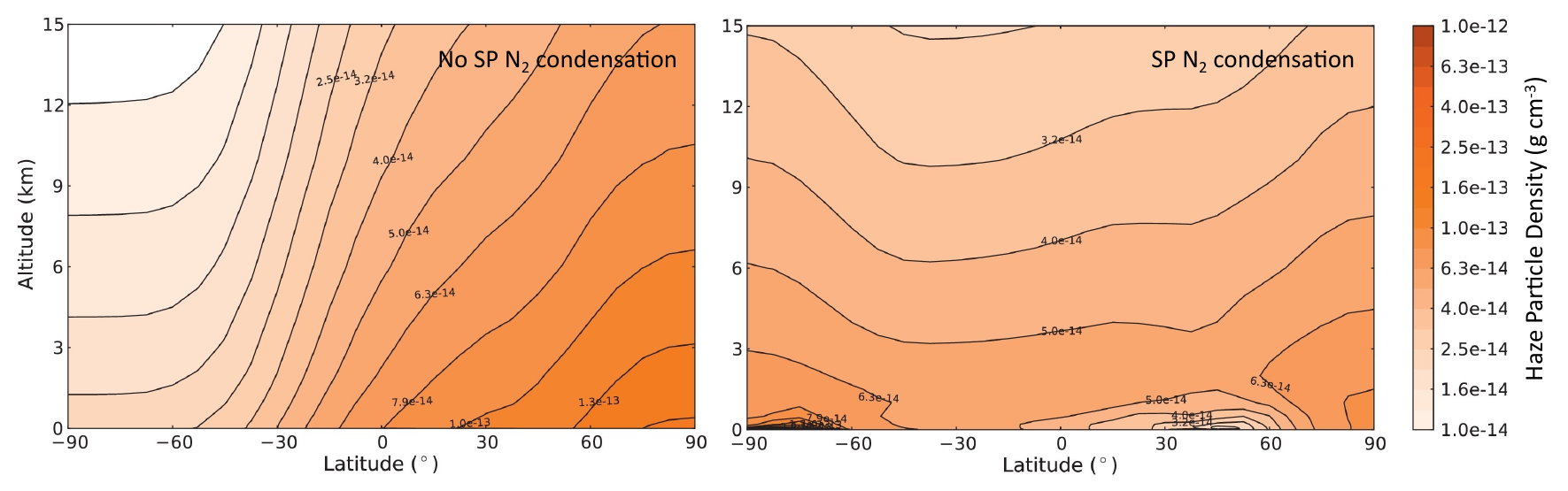}
\end{center}
\caption{Zonal mean Pluto haze particle density without (left) and with (right) \ce{N2} condensation at the south pole, as computed by a 3D model. Figure taken from \cite{Bertrand2017Icar..287...72B}.}
\label{fig:plutodynamics}
\end{figure}

In addition to sedimentation, aerosol particles are also transported by atmospheric circulation. Unfortunately, only one study \cite{Bertrand2017Icar..287...72B} has investigated the impact of circulation on Pluto's hazes, and none have focused on similar processes on Triton. \cite{Bertrand2017Icar..287...72B} simulated the formation, sedimentation, and horizontal advection of haze particles at the time of the New Horizons flyby and showed that the haze distribution is strongly tied to whether \ce{N2} is condensing out on the (unobserved) south pole. Without \ce{N2} condensation, the meridional flow is weak and hazes do not experience significant horizontal transport, resulting in a concentration of hazes in the northern (summer) hemisphere where the Lyman-$\alpha$ flux is the highest. In contrast, \ce{N2} condensation strengthens north-to-south meridional winds due to conveyance of \ce{N2} from Sputnik Planitia to the south pole, resulting in a more homogeneous haze latitudinal distribution. Hazes would also be depleted near the surface over Sputnik Planitia due to the upward branch of the flow, and enriched over the south pole due to the downward branch (Figure \ref{fig:plutodynamics}). Current observations cannot discriminate between these two scenarios \cite{Cheng2017Icar..290..112C,Chen2021Icar..35613976C}.

The observed layering of hazes in Pluto's atmosphere \cite{Gladstone2016Sci...351.8866G,Cheng2017Icar..290..112C,Jacobs2021Icar..35613825J} may be caused by atmospheric waves, which have been observed as temperature perturbations during groundbased stellar occultations, e.g. \cite{Pasachoff2005AJ....129.1718P,Young2008AJ....136.1757Y,Person2008AJ....136.1510P}. These waves could be caused by thermal tides arising from \ce{N2} sublimation \cite{Toigo2015Icar..254..306T,French2015Icar..246..247F} and/or gravity waves arising from orographic forcing \cite{Gladstone2016Sci...351.8866G,Cheng2017Icar..290..112C}. Such waves could act to compactify and rarefy haze distributions, generating the observed layers. However, neither explanation is perfect: the thermal tide hypothesis does not explain how haze particles are transported by the waves and assumes zero mean flow while the orographic forcing hypothesis involves topography that may be too localized to explain the global haze layers \cite{Jacobs2021Icar..35613825J}. Follow up work is needed to model these processes more rigorously.

\subsubsection{Haze Particle Growth}

Haze particles can grow as they are transported in the atmosphere. The primary pathway for growth of involatile haze particles is coagulation, when particles collide and stick together. For the submicron particles observed in Pluto's and Triton's atmospheres, thermal Brownian motion is the main driver of relative velocities and collisions. While it is typically assumed that every particle collision leads to sticking, microphysical studies of Titan's hazes have shown that haze particle charging via interactions with ions and electrons may prevent sticking, thereby regulating the coagulation rate and haze particle size distribution e.g. \cite{Lavvas+10}. The impact of particle charge is typically parameterized by a charge-to-radius ratio.

Growth of haze particles via coagulation can lead to non-spherical particle shapes, which drastically alters their aerodynamical and optical properties. Non-spherical haze particles have typically been treated as fractal aggregates composed of smaller spherical monomers. The shape of a fractal aggregate is characterized by the fractal dimension $D_{\rm f}$, which is related to the number of monomers in an aggregate $N_{\rm mon}$, monomer radius $r_{\rm mon}$, and aggregate characteristic radius $r_{\rm agg}$ (the radius of a sphere that has the same gyration radius) via
\begin{equation}\label{eq:Df}
    N_{\rm mon}=k_{\rm 0}\left( \frac{r_{\rm agg}}{r_{\rm mon}}\right)^{D_{\rm f}},
\end{equation}
where $k_{\rm 0}$ is an order unity prefactor.
Particles with $D_{\rm f}=1$ are chain-like aggregates, while a $D_{\rm f}=2$ particle possesses a mass that is proportional to its cross sectional area. Equation \eqref{eq:Df} indicates that the internal density of aggregates decreases with addition of monomers, while the terminal velocity of the aggregate is the same as that of its monomers when $D_{\rm f}\sim2$, which is typical of aggregate formation via ballistic cluster-cluster collisions \cite{Cabane+93}.

\begin{figure}[t!]
\begin{center}
\includegraphics[width=1.0\textwidth]{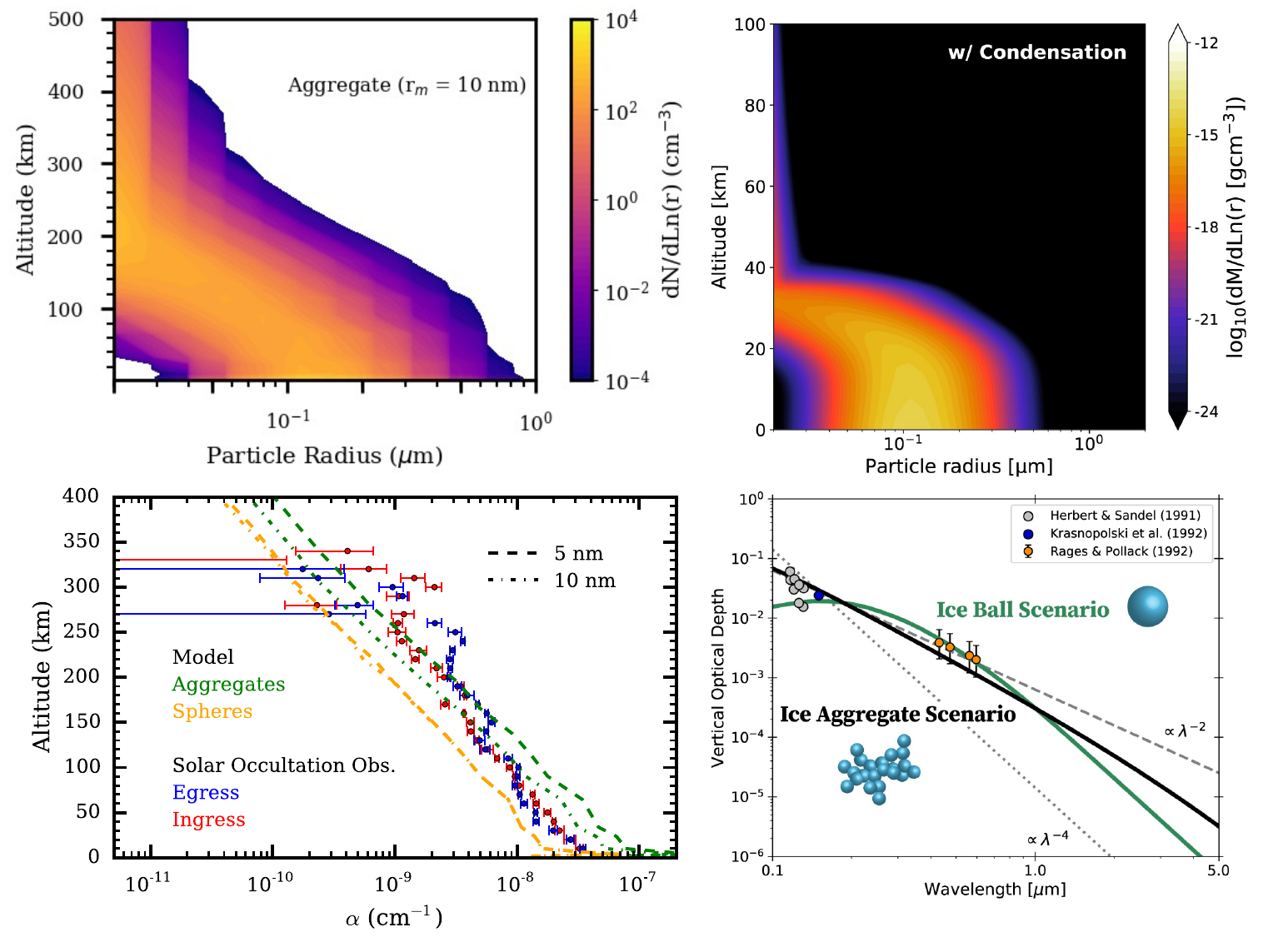}
\end{center}
\vspace{-5mm}
\caption{Number densities of aerosol particles as a function of altitude and particle radius for Pluto hazes, assuming aggregates with 10 nm monomers (top left) and Triton hazes and ice clouds (top right) from the microphysical models of \cite{Gao2017Icar..287..116G} and \cite{Ohno2021ApJ...912...37O}, respectively. \textbf{Bottom left}: Comparison of extinction coefficient $\alpha$ of spherical (orange) and aggregate (green) haze particles with 5 nm (dashed) and 10 nm monomers (dash dot; the 10 nm aggregate model is the same as that presented in the top left panel) with the observed ingress (red) and egress (blue) solar occultation observations at Pluto \cite{Gladstone2016Sci...351.8866G}. \textbf{Bottom right}: Comparison of the vertical optical depth of the ice balls (green line; same model as that plotted top right) and ice aggregates (black line) models for Triton aerosols with the observed optical depths from \cite{Herbert1991JGR....9619241H} (gray), \cite{Krasnopolsky1992JGR....9711695K} (blue), and \cite{Rages1992Icar...99..289R} (orange). All figures taken from \cite{Gao2017Icar..287..116G} and \cite{Ohno2021ApJ...912...37O}.}
\label{fig:microphysics}
\end{figure}

Haze microphysical models of Pluto and Triton typically simulate the vertical transport and growth (by coagulation) of haze particles initially produced at the pressure level of \ce{CH4} photolysis, with the production rate as a free parameter. For Pluto, microphysical models typically require haze production rates on par with the methane photolysis rate due to solar and local interstellar medium-scattered UV (a few times $10^{-14}$ g cm$^{-2}$ s$^{-1}$) to match the observed stellar occultation light curves \cite{Stansberry1989GeoRL..16.1221S} and UV extinction from New Horizons \cite{Gao2017Icar..287..116G}. The latter study also found that aggregate particles with 10 nm monomers and $D_{\rm f}=2$, similar to haze particles on Titan \cite{Rannou+97}, are preferred over compact spherical particles (Figure \ref{fig:microphysics}). \cite{Gao2017Icar..287..116G} further found that reproducing the inferred near-surface particle size of $\sim$0.1 $\mu$m observed by LORRI \cite{Gladstone2016Sci...351.8866G} necessitated particle charging, with a charge-to-radius ratio of $\sim30~{\rm e^-~{\mu}m^{-1}}$, about twice that of Titan \cite{Lavvas+10}. However, subsequent comparisons of the models of \cite{Gao2017Icar..287..116G} to MVIC data suggested a higher value of $\sim60~{\rm e^-~{\mu}m^{-1}}$ \cite{Kutsop2021PSJ.....2...91K}. For Triton, similar modeling has been conducted by \cite{Ohno2021ApJ...912...37O}, who in addition explicitly simulated the evolution of aggregate volume through different-sized aggregate collisions in each mass bin. They found that haze particles on Triton can grow into aggregates with $D_{\rm f}\sim1.8$--$2.2$, though the available observations are insufficient for constraining the actual fractal dimension of Triton aerosols and their charge-to-radius ratios. Critically, \cite{Ohno2021ApJ...912...37O} found that hazes could not simultaneously explain both the observed UV extinction coefficient and visible scattered-light intensity due to absorption by the haze, and that brighter material, such as hydrocarbon ices, must be present. 

\subsubsection{Cloud Formation}

The low temperatures of Triton's and Pluto's atmospheres likely lead to ice cloud formation. Cloud particles form through nucleation, when gas molecules cluster together freely (homogeneous nucleation) or around a cloud condensation nucleus (CCN; heterogeneous nucleation) to form a volatile particle. In the latter case, the CCN is typically another type of particle in the atmosphere. The nucleation rate is dependent on the saturation vapor pressure and the material properties of the condensate, including its surface energy and molecular weight; in the case of heterogeneous nucleation, the rate also depends on the number density, size, and material properties of the CCN, in particular the desorption energy of condensate molecules on the CCN surface and the contact angle between the condensate and the CCN. In general, high surface energies and mean molecular weights, low CCN number densities, small CCN sizes, low desorption energies, and large contact angles lead to lower nucleation rates for a fixed temperature and condensate vapor supersaturation \cite{pruppacher1978}. The inclusion of charged particles can increase the nucleation rate through the introduction of electric potential energy, lowering the energy barrier to nucleation \cite{Moses+92_nucleation}. 

Following nucleation, cloud particles can grow through condensation. In the gas kinetics regime, the condensation rate depends on the abundance and thermal velocity of the condensate vapor, as well as the cloud particle size through the Kelvin curvature effect. The Kelvin effect is the increase in the saturation vapor pressure over a curved surface as compared to that over a flat surface owing to the weaker binding energy of molecules on the former \cite{pruppacher1978}. As such, smaller particles tend to grow slower and evaporate faster. 

Ice clouds as an explanation for the aerosol structures on Triton was a popular hypothesis following the Voyager 2 encounter, owing to their high albedo and supporting evidence from photochemical models \cite{Smith1989Sci...246.1422S,Strobel1990GeoRL..17.1729S,Pollack1990Sci...250..440P,Hillier1991JGR....9619203H,Rages1992Icar...99..289R,Strobel&Summers95_book}. In particular, \cite{Strobel1990GeoRL..17.1729S,Strobel&Summers95_book,Krasnopolsky1995JGR...10021271K} showed that \ce{CH4} photolysis yielded sufficient \ce{C2} hydrocarbons that condensed out in the lower 30 km of Triton's atmosphere (column production rate of $\sim4-8\times10^{-15}$ g cm$^{-2}$ s$^{-1}$) to match the observed optical depth of the global aerosol structure. Meanwhile, the discrete aerosol structures below $\sim$8 km were assumed to be \ce{N2} ice clouds, as they were mostly detected over the southern \ce{N2} polar cap where they could form through large scale upwelling from surface \ce{N2} sublimation and/or plume activity, with \ce{N2} vapor possibly nucleating on the \ce{C2} hydrocarbon ice particles \cite{Pollack1990Sci...250..440P,Rages1992Icar...99..289R}. More recently, \cite{Ohno2021ApJ...912...37O} considered condensation of \ce{C2H4} onto haze CCN and found that the resulting ice cloud could explain both the observed UV extinction coefficient and visible scattered-light intensity assuming a \ce{C2H4} production rate consistent with the photochemical production rate of higher order hydrocarbons computed by \cite{Strobel1990GeoRL..17.1729S} (Figure \ref{fig:microphysics}). The ice cloud is present primarily below 30 km, matching the vertical distribution of Triton's global aerosol layer. Under this scenario, haze CCN is produced at $>$250 km with a rate of only 6 $\times$ 10$^{-17}$ g cm$^{-2}$ s$^{-1}$, resulting in little observable haze above 30 km. These results were corroborated by \cite{Lavvas2021NatAs...5..289L}, who further found through tracking the evolution of $D_{\rm f}$ from relative contributions of ice condensation and coagulation that $D_{\rm f}$ = 3 above 50 km and transitions to 2 below 25 km as coagulation becomes more important.

Ice cloud nucleation onto haze CCN in Pluto's middle atmosphere may explain the depletion of gaseous \ce{HCN} and \ce{C2}-hydrocarbons below 400 km. Using a photochemical model, \cite{Wong2017Icar..287..110W} showed that the shapes of their observed mixing ratio profiles can be reproduced if they condensed onto haze CCN, though the \ce{C2H4} saturation vapor pressure may need to be lower than that extrapolated from laboratory measurements, or else atmospheric mixing may need to be stronger and/or ion chemistry may need to be considered \cite{Krasnopolsky2020Icar..33513374K}. In contrast, \cite{Luspay-Kuti2017MNRAS.472..104L,Mandt2017MNRAS.472..118M}  considered adsorption of gas molecules on haze particles and found similar magnitudes of gas depletion. Adsorption is more dependent on the surface chemistry of haze particles than saturation vapor pressures, with \cite{Luspay-Kuti2017MNRAS.472..104L} finding that Pluto's hazes must become less ``sticky'' to gas molecules as they age due to chemical evolution, which is similar to what is seen for Titan's hazes \cite{Dimitrov2002Icar..156..530D}. \cite{Lavvas2021NatAs...5..289L} also found that adsorption of \ce{C2}-hydrocarbons is possible, but on ice clouds made of \ce{HCN}, \ce{C3H4}, \ce{C4H2}, and \ce{C6H6} instead of haze CCN, as they found condensation of these ices to be abundant in their model below 400 km. \cite{Lavvas2021NatAs...5..289L} further found that ice clouds could make up a significant fraction of the mass of Pluto aerosols, and that $D_{\rm f}$ = 3 above 400 km and 2 below 300 km, with a smooth transition in between as sedimenting spherical particles coagulate into aggregates. 

Gas condensation also impacts Pluto's upper and near-surface atmosphere. At high altitudes where haze CCN is absent, nucleation rates of ice particles may be extremely slow \cite{Rannou2018Icar..312...36R}, potentially explaining the large ($>$10$^6$) \ce{HCN} supersaturation observed by ALMA \cite{Lellouch2017Icar..286..289L}\footnote{In comparison, supersaturation of water vapor on Earth almost never exceeds a few \% due to the ample abundance of CCN \cite{pruppacher1978}.}. Meanwhile, \cite{Stern2017AJ....154...43S} simulated the condensation of hydrocarbon and nitrile ice clouds within 15 km of Pluto's surface where  temperatures reach 40 K and found relatively low cloud optical depths ($\tau\sim$10$^{-4}$), though small ($\sim$1 m s$^{-1}$) updrafts could promote much more optically thick ($\tau\sim$0.1) clouds. Such cloud formation could lead to a sharp increase in $D_{\rm f}$, as shown by \cite{Lavvas2021NatAs...5..289L}. It should be noted, however, that the microphysical parameters used in all of the aforementioned studies, such as the saturation vapor pressure, surface energy, and contact angles between the condensing gases and haze CCN are largely unknown and/or uncertain due to a lack of laboratory measurements at the relevant temperatures. 

\subsection{Summary}\label{sec:synthesis}

The photochemical destruction of \ce{N2} and \ce{CH4} in Triton's and Pluto's atmospheres likely serves as the origins of the clouds and hazes therein. Such photochemical processes produce an array of simple hydrocarbon and nitrile molecules that can react further to form massive charged ions and nm-sized organic haze particles. Alternatively, the simple molecules can condense out and form highly reflective ice clouds that likely nucleate on haze CCN. Models of Triton's and Pluto's aerosols predict that the differences in the aerosol composition and distributions of the two bodies are likely caused by the gaseous methane abundance near their surfaces: the relatively high abundance of methane in Pluto's atmosphere allows methane to survive to hundreds of km altitude, where its photolysis leads to the formation of complex haze particles; in contrast, the lower abundance of methane in Triton's atmosphere results in its photolysis within a few tens of km of its surface, leading to the production of hydrocarbon molecules that condense to form ice clouds. In both Triton and Pluto's atmospheres, aerosol particles sediment from their source towards the surface, with little impact from diffusional processes, though global circulation may be important on Pluto. Both spherical and fractal aggregate particles are likely present, and they may transition from one form to another depending on whether condensation or coagulation is dominant. Finally, it is uncertain whether condensation or sticking/adsorption is more important in Pluto's atmosphere due to uncertainties surrounding the microphysical parameters of the relevant gaseous and CCN species such as their saturation vapor pressures and surface energies.

\section{Laboratory Investigations of Triton and Pluto Aerosols}\label{sec:lab}

\begin{figure}[t]
\centering
\includegraphics[width=1.0\textwidth]{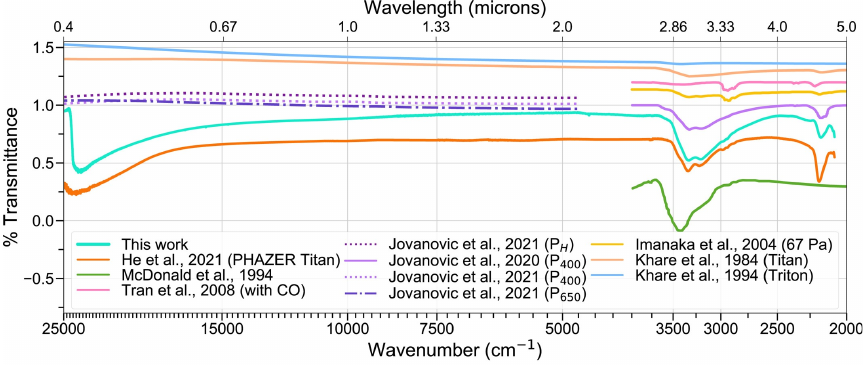}
\caption{Transmittance spectra of organic haze analogs produced in laboratory experiments from mixtures of \ce{N2} and \ce{CH4} (and \ce{CO} in some cases) replicating the atmospheres of Titan (orange, pink, yellow, and salmon), Pluto (light/dark purple dotted and dash-dot), and Triton (cyan, green, and light blue). See Table 3 of \cite{Moran2022JGRE..12706984M} for details on the individual studies. Figure taken from \cite{Moran2022JGRE..12706984M} (``This work'' in the figure).}
\label{fig:tholin_spectrum}
\end{figure}

Laboratory experiments are essential for constraining the material and optical properties of aerosols necessary for modeling aerosol formation and evolution and interpreting observations. To this end, a number of studies have attempted to synthesize photochemical haze analogs by subjecting gas mixtures replicating the compositions of Pluto's and Triton's atmospheres to an energy source. For example, both \cite{Thompson1989GeoRL..16..981T} and \cite{McDonald1994Icar..108..137M} analyzed Triton haze analogs produced from exposing a \ce{N2}/\ce{CH4}($0.1\%$) gas mixture to cold plasma and found remarkable differences in their C/N ratio and spectral features compared to Titan haze analogs, indicating that differences in \ce{N2}/\ce{CH4} ratios led to different reaction pathways in haze production, and that electrons coming from Neptune's magnetosphere likely promote the production of absorbing organic hazes. More recently, \cite{Moran2022JGRE..12706984M} performed a cold plasma experiment on a N$_2$/CH$_4$(0.2\%)/CO(0.5\%) gas mixture at $90~{\rm K}$ and found that the addition of CO results in nitrogen-rich (rather than carbon-rich) Triton haze analogs that incorporate oxygen at $\sim$10 wt\%, indicating that the relative abundance of CO also plays key roles in controlling the properties of organic hazes. Meanwhile, \cite{Jovanovic2020Icar..34613774J,Jovanovic2021Icar..36214398J} investigated the altitude dependence of haze composition on Pluto by considering a \ce{N2}/\ce{CH4}/\ce{CO} gas mixture where the \ce{CO} abundance is fixed to 500 ppm and the \ce{CH4} abundance is allowed to vary from 0.5 to 5\% to simulate high altitude diffusive separation. They found that more N and O atoms were incorporated in the sample with lower \ce{CH4} abundances, showing that haze composition likely varies with altitude. They also found that the Pluto haze analogs were less absorbing at optical wavelengths compared to Titan haze analogs, which are produced without \ce{CO} gas (Figure \ref{fig:tholin_spectrum}). Finally, \cite{He2017ApJ...841L..31H} considered a \ce{N2}/\ce{CH4}/\ce{CO} gas mixture with varying \ce{CO} abundances that is applicable to Titan, Triton, and Pluto, and found an increase in the density and oxygen content and a decrease in the saturation of the haze analogs with increasing \ce{CO}.

To aid our understanding of ice cloud formation in Triton's and Pluto's atmosphere, several studies have measured and compiled the saturation vapor pressures and surface energies of relevant hydrocarbon and nitrile ices \cite{Moses+92_nucleation,Fray2009P&SS...57.2053F,Yu+22_material_properties}. However, the temperature ranges for which the measurements are valid often do not contain the low temperatures of Triton and Pluto, and thus caution is required when extrapolating the empirical saturation vapor pressure and surface energy expressions. Measurements of other microphysical parameters, such as the desorption energies and contact angles between condensates and CCN are relatively rare, and confined mostly to methane and ethane condensates and haze CCN \cite{Curtis2008Icar..195..792C,Rannou2019A&A...631A.151R}. 

\section{Triton and Pluto Aerosols: Unknowns and Outlook}\label{sec:conclusions}

Despite similarities in atmospheric composition and surface pressures and temperatures, the aerosols of Triton and Pluto are distinct from each other (Table \ref{table:compare}). Triton hosts a global aerosol layer likely composed of organic ices that is confined to the lower few tens of km of its atmosphere, with discrete clouds below 10 km, while Pluto's aerosols are seemingly dominated by organic hazes extending up hundreds of km with multiple fine layers, though organic ice adsorption and/or condensation also likely play a role. This difference has been hypothesized as being driven by the difference in \ce{CH4} abundance on the two worlds: the lower \ce{CH4} abundance on Triton leads to \ce{CH4} and \ce{N2} photochemistry in the lower few tens of km, leading to production and condensation of simple hydrocarbon and nitrile ices. Meanwhile, the higher \ce{CH4} abundance on Pluto allows it to survive into the upper atmosphere, resulting in the production of organic haze particles at hundreds of km altitude from a rich network of neutral and ion reactions; these particles then sediment downwards and could act as nucleation centers and/or adsorbing surfaces for gaseous hydrocarbon and nitrile species. 

Groundbased stellar occultations, the Voyager 2 and New Horizon flybys, and lessons learned from Cassini observations of Titan have increased our knowledge of Triton's and Pluto's aerosols considerably, but many vital unknowns remain. Due to the limitations in wavelength and phase angle of the existing observations, the composition, and therefore formation mechanism of Triton's and Pluto's aerosols remain uncertain. While parallels could be drawn between the organic hazes and ice clouds of Titan and those on Triton and Pluto, the different atmospheric compositions (i.e. higher \ce{CO} on the latter bodies), energy sources, and ionospheric environments could result in significant differences in the formation and evolution of the aerosols of the three worlds. In addition, the impact of aerosols on the atmospheric radiative transfer and dynamics of Triton and Pluto requires further study. For example, \cite{Zhang2017Natur.551..352Z} explained the observed low escape rate of Pluto's atmosphere by showing that Pluto's aerosols could significantly cool its upper atmosphere. This would indicate a fundamental connection between atmospheric and surface volatile evolution, photochemistry, and aerosol formation in Pluto's atmosphere that is likely modulated by Pluto's extensive seasonal cycles caused by its eccentric orbit \cite{Johnson2021Icar..35614070J}. Similar connections between Triton's aerosols and the rest of Triton's atmosphere may also exist and therefore should be investigated.

\begin{table*}[t]
  \caption{Summary of Triton and Pluto aerosol properties. See ${\S}$\ref{sec:obs} and ${\S}$\ref{sec:theory} for details.}\label{table:compare}
  \centering
  \begin{threeparttable}
  \begin{tabular}{l| l  l} \hline\hline
      & Triton & Pluto\\ \hline \hline
    \multirow{2}{*}{Spatial distribution} & global + discrete structures, & global, layered,\\ 
     & up to 30 km \cite{Smith1989Sci...246.1422S,Pollack1990Sci...250..440P}& up to 500 km \cite{Gladstone2016Sci...351.8866G} \\ \hline
    Production rate & 2--8$\times$10$^{-15}$ g cm$^{-2}$ s$^{-1}$ \cite{Ohno2021ApJ...912...37O}& $\sim$10$^{-14}$ g cm$^{-2}$ s$^{-1}$ \cite{Gao2017Icar..287..116G}\\ \hline
    \multirow{2}{*}{Composition} & N$_2$ ice (discrete) \cite{Pollack1990Sci...250..440P} & \multirow{2}{*}{organic haze + ices \cite{Gao2017Icar..287..116G,Lavvas2021NatAs...5..289L}}\\
    & organic ices (global)\cite{Ohno2021ApJ...912...37O,Lavvas2021NatAs...5..289L}& \\\hline
    \multirow{2}{*}{Size} & $\sim$0.2-1.5 $\mu$m (discrete),\cite{Pollack1990Sci...250..440P,Rages1992Icar...99..289R} & \multirow{2}{*}{$\sim$0.1-1 $\mu$m \cite{Fan2022NatCo..13..240F}}\\
    & $\sim$0.15 $\mu$m (global) \cite{Krasnopolsky1992JGR....9711695K,Rages1992Icar...99..289R}& \\ \hline
    Shape & unconstrained \cite{Ohno2021ApJ...912...37O}& spheres + aggregates\cite{Fan2022NatCo..13..240F}\\ \hline
    Charge & unconstrained \cite{Ohno2021ApJ...912...37O}& 30-60 e$^-$ $\mu$m$^{-1}$ \cite{Gao2017Icar..287..116G,Kutsop2021PSJ.....2...91K}\\ \hline
    \multirow{2}{*}{Optical depth} & $\sim$0.03 (discrete), \cite{Pollack1990Sci...250..440P} & \multirow{2}{*}{$\sim$0.01 \cite{Gladstone2016Sci...351.8866G}}\\
    &  $\sim$0.003 (global) \cite{Rages1992Icar...99..289R} & \\ \hline
    Single scattering albedo & 0.99 \cite{Hillier1990Sci...250..419H,Hillier1991JGR....9619203H}& 0.9-0.95\cite{Hillier2021PSJ.....2...11H}\\
    %Column mass &  & $\sim$10$^{-7}$ g cm$^{-2}$& \cite{Gladstone2016Sci...351.8866G}\\
    \hline \hline
  \end{tabular}
  %\begin{tablenotes}
       % \item[a] We combine the two types of aerosols in Triton's atmosphere here for brevity. See ${\S}$\ref{sec:tritonvoy2} for how the properties of the global aerosol layer and the discrete structures differ.
    %\end{tablenotes}
\end{threeparttable}
\end{table*}

Furthering our understanding of Triton's and Pluto's aerosols requires synergy between observations, modeling, and laboratory studies. A dedicated orbiter at Triton and/or Pluto would greatly increase our knowledge of their aerosols by observing their atmospheres at a wide range of phase angles and wavelengths, and by monitoring them for temporal and spatial variations in aerosol distributions and optical properties. In particular, spectral features associated with solid organic material are abundant at longer, mid-infrared wavelengths \cite{Zhang2017Natur.551..352Z}. Meanwhile, laboratory measurements of key microphysical parameters at Triton and Pluto temperatures, such as the saturation vapor pressure, surface energy, desorption energy, and contact angles of organic ices and haze analogs would be essential for more accurate modeling of ice cloud formation and how hydrocarbon and nitrile species interact with organic haze particles. Further laboratory work on haze formation is also needed, particularly in how they initially form from gaseous species and how they grow through condensation and/or coagulation into spheres or aggregates. In addition, uncertainties in the aerosols' optical properties remain a major bottleneck in interpreting observations and thus more measurements are needed. At the same time, modeling efforts that take into account the laboratory measurements and additionally focus on how hazes interact with the 3D radiative environment and dynamics of the atmosphere, along with how they evolve on seasonal timescales would also greatly aid in understanding current and future observations. Finally, given that the differences between Triton's and Pluto's aerosols likely hinge on the surface \ce{CH4} abundance, observing and modeling these surfaces and the near-surface atmosphere through time would help us better understand whether Triton's and Pluto's aerosols are fundamentally different or they are simply different points on the same evolutionary path. 

\newpage

\bibliographystyle{plain}
\bibliography{References}

\end{document}